\documentstyle[12pt]{article}
\begin{document}
\textwidth=16cm
\textheight=21cm

\title
{{\rightline{\small{TIFR-TH/95-60}}}
{\rightline{}}
DEGREES OF FREEDOM IN TWO DIMENSIONAL STRING THEORY
\footnote{Based on talks at Spring Workshop on String Theory and
Quantum Gravity at ICTP, Trieste, March 1995 and at VIIth Regional
Conference on Mathematical Physics, Bandar-Anzali, October 1995.}}
\author{Sumit R. Das \thanks{E-mail: das@theory.tifr.res.in}\\
{\small{{\em Tata Institute of Fundamental Research}}} \\
{\small{{\em Homi Bhabha Road , Bombay 400005, INDIA.}}}}
\date{}
\maketitle

\begin{abstract}
We discuss two issues regarding the question of degrees of freedom in
two dimensional string theory. The first issue relates to the
classical limit of quantum string theory.  In the classical theory one
requires an infinite number of fields in addition to the collective
field to describe ``folds'' on the fermi surface. We argue that in the
quantum theory these are not additional degrees of freedom. Rather
they represent quantum dispersions of the collective field which are
{\em not} suppressed when $\hbar \rightarrow 0$ whenever a fold is
present, thus leading to a nontrivial classical limit.  The second
issue relates to the ultraviolet properties of the geometric
entropy. We argue that the geometric entropy is finite in the
ultraviolet due to {\em nonperturbative} effects. This indicates that
the true degrees of freedom of the two dimensional string at high
energies is much smaller than what one naively expects.
\end{abstract}

\section{Introduction}

Recent developments seem to indicate that the true degrees of freedom
in string theory are quite different from what is expected from
perturbation theory. In fact we already know this from our experience
with noncritical strings : the fundamental formulation of two
dimensional string theory is in terms of nonrelativistic fermions via
matrix models and a stringy description in terms of the massless
tachyon emerges as a low energy perturbative picture \cite{DONE}. For
higher dimensional strings we do not have a nonperturbative
formulation, but the remarkable duality properties indicate that
string theory has to be formulated in terms of some yet unknown
entities which are not the strings of perturbation theory.

In this talk I will summarize some results which clarify the nature of
degrees of freedom of the two dimensional string. This is a good
laboratory since some quantities can be calculated nonperturbatively.
, even though the space-time interpretation of the
theory is rather involved and not yet fully understood. The first set
of results, obtained in collaboration with S. Mathur \cite{DASMATHUR}
indicate that the classical limit of the theory is rather
nontrivial and requires more degrees
of freedom than the quantum theory itself.  The second set of results
\cite{DASENT} indicate that a suitably defined geometric entropy (or
entropy of entanglement), which provides a measure of the degrees of
freedom, is ultraviolet finite due to essentially {\em nonperturbative}
effects This is relevant to the question of black hole entropy in
string theory.

\section{Folds and the Classical limit}

Two dimensional string theory is described by $N$ mutually
nointeracting nonrelativistic fermions in $1+1$ dimensions
in an external inverted harmonic oscillator potential with
the second quantized action in terms of fermi fields $\psi (x,t)$
\begin{equation}
A_f = \lambda\int dt \int dx~
\psi^\dagger[-{1 \over 2\lambda^2}\partial_x^2 +\mu_c - {1\over 2} x^2]
\psi
\label{eq:one}
\end{equation}
and $\lambda = {N \over g}$ and $g$ is the quartic coupling
in the underlying matrix model. If $\mu$ denotes the fermi
level of the single particle hamiltonian in (\ref{eq:one}),
the continuum string theory is described by the double
scaling limit $N \rightarrow \infty$ and $\mu \rightarrow \mu_c$
with $\kappa = \lambda(\mu_c - \mu) = {\rm fixed}$.

\subsection{Classical Dynamics}

Let us first consider the classical dynamics of the fermion
system in a manner which is independent of the form of the
hamiltonian. The classical state is specified by a distribution
function $u(x,p,t)$ in phase space
- in fact $u (x,p,t) = {1 \over 2\pi \hbar}$
in some region bounded by the fermi surface and zero elsewhere.
The fermi sea may be conveniently paramterized in terms of
two functions $\beta_\pm$ and an infinite set of additional
fields $w_{n,\pm}$ \cite{AVJEV}
\begin{eqnarray}
\int dp~u(x,p,t) & = & {1 \over 2\pi \hbar}[\beta_+ - \beta_-] \nonumber \\
\int dp~p~u(x,p,t) & = & {1 \over 2\pi \hbar}[{1\over 2}(\beta_+^2
- \beta_-^2) \nonumber \\
& & + (w_{+1}-w_{-1})] \nonumber \\
\int dp~p^2~u(x,p,t) & = & {1 \over 2\pi \hbar}[{1\over 3}(\beta_+^3
- \beta_-^3) \nonumber \\
& & + (\beta_+w_{+1} - \beta_-w_{-1}) \nonumber \\
& & + (w_{+2}-w_{-2})]
\label{eq:two}
\end{eqnarray}
and so on. Let a line of constant $x$ in  phase space
cut the upper edge of the fermi sea at points $p_i(x,t)$
and the lower edges of the fermi sea at $q_i(x,t)$ where
$ (i = 1,2,\cdots i_m(x,t))$. Then
\begin{equation}
\int dp~p^n~u(x,p,t) =
\sum_{i=1}^{i_m} {[p_i^{n+1} - q_i^{n+1}] \over 2\pi \hbar(n+1)}
\label{eq:eleven}
\end{equation}
$i_m(x,t)- 1$ denotes the number of ``folds'' at
the point $x$ at time $t$. In the absence of folds, one can
set $\beta_+(x,t) = p_1 (x,t)$ and
$\beta_-(x,t) = q_1 (x,t)$ which then implies that all the
$w_{\pm , n} = 0$. This is the standard bosonization in terms of
the collective field theory. As emphasized in \cite{POLF}
only in the absence of folds  is the state  described by
a scalar field $\eta(x,t)$ and its momentum conjugate $\Pi_\eta
(x,t)$,  which are related to $\beta_\pm$ by $\beta_\pm
= \Pi_\eta \pm \partial_x \eta$.

In the deep asymptotic region, the string theory massless tachyon is a
nonlocal transform of $\eta (x,t)$ \cite{POLTRANS}.  In the presence
of folds we do not know how to extract the string theory space-time
from the matrix model, since the $w_{\pm,n} \ne 0$ while in the far
asymptotic region the collective field configurations alone exhaust
the possible configurations of the string theory tachyon
field \footnote{Away from the asymptotic region higher moments of
$u(x,p,t)$ are required for determining the string theory tachyon. See
Dhar et. al. in \cite{POLTRANS}.}.

As shown in \cite{AVJEV} the Poisson bracket algebra for $\beta_\pm$
is that of a chiral (antichiral) boson, while the $w_{n,\pm}$
commute with the $\beta_\pm$ and themselves satisfy a
$w_{\infty}$ algebra, for the $\pm$ components separately.  In the
following we will omit the $\pm$ subscript.

We will also concentrate on {\em free} fermions.
A background potential can be easily incorporated and is not
essential to the main point.

For free relativistic right moving fermions one has
a hamiltonian $H = \int dx \int dp~p~u(x,p,t)$. The
Poisson brackets lead to the equations of motion
\begin{equation}
\partial_t\beta = -\partial_x \beta~~~~~\partial_t w_m
= -\partial_x w_m
\label{eq:fifteen}
\end{equation}
For free nonrelativistic fermions one has $H = \int dx
\int dp~{1\over 2}p^2~u(x,p,t)$
and using the Poisson brackets one gets the evolution equations
\begin{eqnarray}
\partial_t \beta  & = & -\beta\partial_x\beta -  \partial_x w_1
\nonumber \\
\partial_t w_m  & = & -2w_m~\partial_x\beta - \beta~
\partial_x w_m - \partial_x w_{m+1}
\label{eq:seventeen}
\end{eqnarray}

The $\beta, w_n$ are expressible in terms of the $(p_i,q_i)$
introduced in (\ref{eq:eleven}). We choose a parametrization
associated to the upper edge of the fermi surface, i.e. we will choose
all the $w_{-,n} = 0$ and
\begin{equation}
\beta_-(x,t) = q_{i_m}~~~~\beta_+ (x,t) = \sum_{i=1}^{i_m} p_i
- \sum_{i=1}^{i_m-1} q_i
\label{eq:twelve}
\end{equation}
For relativistic fermions the fermi sea has no lower edge and
one has to set all $\beta_- = w_{-n} = 0$.  Using (\ref{eq:twelve}),
(\ref{eq:two}) and (\ref{eq:eleven}) we can now easily calculate all
the $w_{\pm,n}$ 's.  Clearly, the $w_n$ 's are independent of $\hbar$.

Each of the
$p_i, q_i$ satisfy an evolution equation determined by the single
particle hamiltonian: $\partial_t p_i = - \partial_x p_i$ for
relatvisitic fermions and $\partial_t p_i = -p_i\partial_x p_i$
for nonrelativistic fermions (and similarly for $q_i$'s). It
may be checked that these equations then imply the corresponding
evolutions for the $\beta$ and $w_m$ 's.

Let the
profile of the fermi surface at $t = 0$ be given by
$p(x,0) = a(x)$. Then at a later time $t$ it is easy to show that
$p(x,t) = a(x-t)$ for relativistic chiral fermions and
$p(x,t) = a(x-p(x,t)~t)$ for nonrelativistic fermions.
If there is a fold, there must be some point where
${dx/dp} = 0$.
Since the profile in a relativistic system
is unchanged in time a fold cannot develop from a profile
with no folds. On the other hand for a nonrelativistic system,
even if we start with a single valued  $a(x)$
one will have  ${dx/dp} =0$
at some point on the fermi surface at some time.
We give the result for the time of fold formation $t_f$ for two
initial profiles:
\begin{eqnarray}
p(x,0) & = & b~e^{-{(x-a)^2 \over c^2}}- (a \rightarrow -a),
{}~~ t_f  \approx  {c e^{{1\over2}}\over b{\sqrt{2}}}
\nonumber \\
p(x,0) & = & k~{\rm Re}(C_k~e^{ikx}),~~~
t_f = {1 \over (\vert C_k \vert k^2)}
\label{eq:twenty}
\end{eqnarray}
where $ a >> c$. Thus fold formation can occur for
pulses of arbitrarily small energy density (and total energy),
provided the width is sufficiently small. In the following we
will only consider low energy pulses so that we can ignore the
presence of the bottom edge of the fermi sea.

It may be noted that the notion of folds depends on a particular
choice of coordinates and momenta in phase space. In fact for
nonrelativistic fermions in an inverted harmonic oscillator potential,
one may perform a transformation to light cone like coordinates $x_\pm
= p \pm x$ in phase space in which the fermions appear {\em
relativistic} and the resulting bosonic fields $\beta_\pm$ would be
{\em free} (see O'Loughlin in \cite{POLTRANS}).  In such a
parametrization a non-folded profile cannot evolve into a
fold. However, in the physical scattering process incoming states are
defined in terms of quanta using $x_-$ as the ``space'' while for the
outgoing state $x_+$ has to be used as ``space''. The classical
scattering matrix is in fact given in terms of a Bogoluibov
transformation Since a nonfolded profile in the $x_-$ parametrization
generically becomes a folded profile in the $x_+$ configuration one
cannot evade the question of fold formation.

It is now clear what happens if we set $w_n = 0$ in the time evolution
equations (\ref{eq:seventeen}). The evolution of $\beta$ proceeds
smoothly till $t = t_f$ at which point $\partial_x \beta$
diverges. The equations cannot be evolved beyond this point and the
collective field theory fails.  The time evolution of $u (p,q,t)$ is
of course unambiguous. The point is that one needs
nonzero values of $w_n$ beyond this time, whose presence render the
classical time evolution of the entire $\beta,w_n$ system completely
well defined. The same phenomenon happens in the presence of
a background potential.

\subsection{The quantum theory}

Let us now turn to the quantum theory.
Since the bottom edge of the fermi sea is irrelevant for the
question of fold formation, we can completely ignore its
presence. Then we have an exact operator bosonization of the
system. Define Schrodinger picture field operators
\begin{equation}
{\hat \psi}(x)  = {1\over \sqrt{L}}\sum_{n=-\infty}^{\infty}
{\hat \psi}_n~e^{i{2\pi
n \over L}x}
\label{eq:five}
\end{equation}
with a similar mode decomposition for $\psi^\dagger (x)$ which
define the modes $\psi^\dagger_m$.
${\hat \psi}_n$, ${\hat \psi
^\dagger}_n$
are annahilation and creation operators for a fermion at level $n$
obeying the standard anticommutation relations
$[{\hat \psi_n}^\dagger,{\hat \psi_m}]_+ ~=~\delta_{n,m}$
and all other anticommutators being zero.
The vacuum is defined by
\footnote{We have chosen phases in (\ref{eq:five})
so that $n = 0$ is the vacuum fermi level}
${\hat \psi}_n |0>  =  0$ for $n > 0$ and
${\hat \psi}^\dagger_n |0>  = 0$ for $n \le 0$.

Let
\begin{equation}
\alpha_{-n} = \sum_{m=-\infty}^\infty :{\hat\psi}^\dagger_{n+m}
 \hat\psi_m :
\label{eq:twentytwo}
\end{equation}
The position space field is defined by a usual fourier transform.
The modes $\alpha_n$ are thus shift operators on fermion levels.
(See e.g. \cite{KACRAINA}.)
The normal ordering is defined according to the vacuum defined
above.
Then it may be verified that
\begin{equation}
[\alpha_n ,\alpha_m ]~=~n\delta_{n+m,0}
\label{eq:twentyfour}
\end{equation}
The inverse correspondence to (\ref{eq:twentytwo}) is also well
known and given e.g.in \cite{KACRAINA}.

In the classical limit, the classical quantities in (\ref{eq:two})
are then related to expectation values of appropriate
operators in some quantum state, $| \xi , t>$
\begin{eqnarray}
\int dp~u & = & < :\psi^\dagger\psi: > = <\alpha (x) >
\nonumber \\
\int dp~p~u  & = & <{i \hbar \over 2}
[ : (\partial_x \psi^\dagger)\psi
-\psi^\dagger (\partial_x \psi ):]> \nonumber \\
& = & 2\pi\hbar <:
{\alpha^2\over 2}:>
\label{eq:twentyeight}
\end{eqnarray}
Similarly, $\int dp~p^n~u \sim <:\alpha^n:>$.  In
(\ref{eq:twentyeight}) $< A > = <\xi,t | A | \xi,t>$.  The second
equalities in (\ref{eq:twentyeight}) follow from the operator
expressions of $\psi$ in terms of $\alpha$ \cite{KACRAINA}.  The
equations (\ref{eq:twentyeight}) demonstrate the main point : {\em
$w_n$ are a measure of the quantum fluctuations in the bosonic
field}. Thus, e.g.  comparing (\ref{eq:two}) and
(\ref{eq:twentyeight}) we have
\begin{equation}
w_1 (x,t) = {(2\pi\hbar)^2\over 2} [<:\alpha^2(x):> -
<\alpha (x)>^2]
\label{eq:twentynine}
\end{equation}
Note that in the normalizations we are using, the fluctuations of
$\alpha$ must be $O({1 \over \hbar^2})$ for $w_n \sim O(1)$.

\subsection{The classical limit of the quantum theory}

Normally the classical limit of a quantum system is obtained by
considering coherent states. In such states, quantum dispersions of
operators are suppressed by powers of $\hbar$. In an interacting
theory, time evolution does not keep the state coherent, but once
again the departures are such that the quantum dispersions are still
suppressed in the $\hbar \rightarrow 0$ limit. Consequently,
expectation values of operators in such coherent states act as
classical dynamical variables and obey the classical equations of
motion.

In our system the coherent states which represent the deformations
of the fermi sea are coherent states of the boson field $\phi(x,t)$.
Consider such a state at $t = 0$ given by
\begin{equation}
|\xi,0 > = \prod_{n=1}^{\infty} e^{{C_n \over \hbar} \alpha_{-n}} | 0 >
\label{eq:thirty}
\end{equation}
It is straightforward to check that
\begin{equation}
<\xi,0| \alpha (x)| \xi,0 > = {2 \over \hbar L}\sum_{n=1}^\infty
{\rm Re}[nC_ne^{2\pi inx/L}]
\label{eq:thirtyone}
\end{equation}
Note that $<\alpha (x)> \sim
{1\over \hbar}$ which is required for $\beta \sim O(1)$
(see  (\ref{eq:twentyeight})). It may be easily shown that
(at $t = 0$)
\begin{equation}
<:\alpha^n (x):> = <\alpha (x)>^n
\label{eq:thirtytwo}
\end{equation}
Thus the $w_n$ vanish at $t=0$, and we have a fermi surface without folds.
In the following we will consider a state with only one nonzero
$C_n$ for $n = {\bar n}$. The classical fluid profile is then
given by (\ref{eq:twenty})
This is sufficient for our purpose.

The time evolution of $n$-point functions of $\alpha (x)$ is
most easily computed in the Heisenberg picture using the free
equations of motion of the fermi fields.
The details of the calculation are given in \cite{DASMATHUR}. To
illustrate our main point it is sufficient to give the result for
the quantity $G(p,t) = <\xi,t|\alpha_{-p} \alpha_p | \xi,t>$
\begin{equation}
G(p,t) = \sum_{s > {p \over {\bar n}}}
({\bar n}s - p)
J_{s}^2(\Phi({\bar n},p))
\label{eq:fortythree}
\end{equation}
where $J_s$ denotes the Bessel function and
\begin{equation}
\Phi(n,p) = {4 |C_n| \over
\hbar}~{\rm sin}~[({2\pi \over L})^2~{\hbar p n t \over 2}]
\label{eq:fortyone}
\end{equation}

It may be seen from (\ref{eq:fortythree}) and (\ref{eq:fortyone})
that when
\begin{equation}
p > p_M = {4 |C_{\bar n}|~{\bar n} \over \hbar}
\label{eq:fournine}
\end{equation}
$G(p)$ vanishes exponentially with increasing $p$ regardless of
the $\hbar \rightarrow 0$ limit. This is because in this case
the index of the Bessel function is {\em always} greater than the
argument and Bessel functions decay exponentially when the ratio of
the index to the argument grows large (see (\ref{eq:fortyfour}) below).

Consider now the classical limit for $p >> {\bar n}$, but $p < p_M$.
The argument of the Bessel fucntion is now smaller than the
index and the dominant
contribution comes from the minimum allowed value of $s$ in
the sum in (\ref{eq:fortythree}) which is
$s_m = {p \over {\bar n}}$. For large $p$
the relevant Bessel function in (\ref{eq:fortythree})
behaves as
\begin{equation}
J_{s_m}(\Phi({\bar n},p)) \sim
{e^{-{p \over {\bar n}}(\beta-\tanh\beta)}\over
(2 {p \over {\bar n}}\pi\tanh\beta)^{(1/2)}}
\label{eq:fortyfour}
\end{equation}
where we have defined
\begin{equation}
{\rm cosh}~\beta = (2  |C_{\bar n}| ({2\pi\over L})^2{\bar n}^2 t)^{-1}
\label{eq:fortyfive}
\end{equation}
So long as $\beta > 0$, $G(p)$ is thus exponentially suppressed
for large $p$. The exponential suppression disappears when
$\beta = 0$ or when
\begin{equation}
t = t_0 = (2  |C_{\bar n}| ({2\pi\over L})^2{\bar n}^2 )^{-1}
\label{eq:fortysix}
\end{equation}
The time $t_0$ is {\em exactly} equal to the time of fold
formation $t_f$ as calculated in equation (\ref{eq:twenty}).

In fact the above results are expected in the absence of folds.
For  the calculation of $G(p,t)$ with
$p >> {\bar n}$ the fermi surface may be considered to be flat in
a small interval in $x$. Since $\alpha_p$ is a lowering operator
for $p > 0$ it is clear that $G(p) = 0$ for such large $p$
since there is no empty level below the fermi surface. However,
if there are folds, there are empty bands separating filled
levels and $G(p)$ can be nonzero. However for $p$ very large
(of $O({1\over \hbar})$ in order to have an order one momentum)
$G(p)$ has to be again zero since $\alpha_p$ would try to
move a fermion into the bulk of the fermi sea.

Indeed for  $t > t_f$ the relevant Bessel function behaves, for large
$p/\bar n$ as
\begin{equation}
J_{{p \over {\bar n}}}
(x)~\sim~
{ \cos({p \over {\bar n}} \tan\beta -n\beta -\pi/4)\over
( {p \over {\bar n}}
\pi\tan\beta)^{(1/2)}}
\label{eq:fortyseven}
\end{equation}
where $\cos \beta = (2  |C_{\bar n}| ({2\pi\over L})^
2{\bar n}^2 t)^{-1}$. At late times
it may be shown that for $p/\bar n>>1$ (but $p << p_M$)
\begin{equation}
G(p) \sim  2  |C_{\bar n}|({2\over L})^2\pi{\bar n}^2 p t~=~p~n_{\rm fold}
\label{eq:fortyeight}
\end{equation}
where
$ n_{{\rm fold}} =
2  |C_{\bar n}|({2\over L})^2\pi{\bar n}^2  t=2\vert p_{max} (x) \vert
{{\bar n} t \over L}$ is the number of folds computed from
the classical motion
of the fermi fluid.

Finally we estimate the quantities $w_n(x,t)$ and see
whether they are nonzero in the classical limit. We will
consider the quantity
\begin{equation}
w_{1,0} (t)  \equiv \int dx w_1 (x,t)
=   2(2\pi\hbar)^2 \sum_{p>0} G(p,t)
\label{eq:fortynine}
\end{equation}
Since $G(p,t)$ decays exponentially for $p > p_M$, we can
effectively put an upper bound on the sum over $p$ at $p_M$.
\begin{enumerate}

\item For $t < t_f$, one has a $G(p)$ which decays
exponentially with $p$ at a rate {\em independent of} $\hbar$ (see
equation (\ref{eq:fortyfour})), and the upper limit of integraion is
irrelevant. Thus in this situation one has $w_{1,0} (t) \sim \hbar^2$
which vanishes in the classical limit.

\item For $t > t_f$ one has $G(p) \sim p$. In this case one
clearly has $w_{1,0}(t) \sim \hbar^2 p_M^2$. Using (\ref{eq:fournine})
one then has $w_{1,0}(t) \sim O(1)$ and survives in the classical
limit.

\end{enumerate}
Thus we see that
the presence of folds in the classical description signifies
quantum fluctuations of the bosonic field which {\em survive
in the} $\hbar \rightarrow  0$ {\em limit.}

While we have demonstrated our result in a simple model, it
is clear from the derivation that our main contention is valid
for the matrix model described in terms of fermions in an
inverted harmonic oscillator potential  - though the details
would be  more complicated. Furthermore the effect of the
lower edge of the fermi sea becomes relevant for states with
high energy. As mentioned the question of folds is independent
of this, but a complete treatment should incorporate this
feature.

\section{Geometric entropy in the 2d string}

Let us now turn to an effect which is a direct result of the presence
of a lower edge of the fermi sea of nonrelativistic fermions. The
physical question relates to the notion of geometric entropy or the
entropy of entaglement and is relevant in black hole physics. The
entropy of black holes has a part which is {\em intrinsic} - the
Hawking-Beckenstein entropy \cite{BEK}. This is a
classical contribution and is usually evaluated by computing the
classical action for euclidean black holes. In addition to this, there
is the quantum correction to the black hole entropy \cite{THOFT} ,
which is the entropy of matter outside the black hole horizon. This,
in turn, is related to the entropy of entanglement \cite{SORKIN} (in a
given quantum state of matter) between the region inside and outside
the black hole.

The large mass limit of a black hole is Rindler space.
In Rindler space the geometric entropy can be obtained as follows.
Let $x$ denote a spatial coordinate in {\em flat} space.
Denote fields in the region $x > 0$
($x < 0$)by $\phi_R (\phi_L)$. In some given quantum state
with a wave functional $\Psi_i[\phi_L,\phi_R]$ the reduced density
matrix (unnormalized) is obtained by integrating out $\phi_L$ is given by
\begin{equation}
\rho[\phi_R,\phi_R'] = \int {\cal D} \phi_L~\Psi_i[\phi_L,\phi_R]
\Psi_i[\phi_L,\phi_R ']
\label{eone}
\end{equation}
Then the geometric entropy is given by
\begin{equation}
S_g = - {\rm tr}[\rho {\rm log} \rho]
\label{etwo}
\end{equation}
For scalar and spinor fields the geometric entropy and the quantum
correction to the Rindler space entropy are exactly the same. For like
gauge fields the two differ by contact terms corresponding to states
exactly on the horizon \cite{KABAT2}.

The geometric entropy is in turn the same as the ordinary thermal
entropy in a space with {\em position-dependent} temperature, the
position dependence being given by the standard Tolman relation in
Rindler space.

A crucial property of the geometric entropy is that it is ultraviolet
divergent in usual field theories. This may be understood from the
above thermodynamic interpretation. For Rindler space the position
dependent temperature is given by $T(x) = {1 \over
\beta(x)} = {1 \over 2\pi x}$. Consider for example a massless scalar
field in $d + 1$ dimensions. The geometric entropy may be calculated
by integrating the entropy density in this position dependent
temperature.
\begin{eqnarray}
S_g & = & \int dx \int {d^dk \over (2\pi)^d}[{\rm log}(1-e^{-2\pi x
 \vert k \vert}) \nonumber \\
& & -{1 \over 1-e^{-2\pi x \vert k \vert}}]
\label{eq:aone}
\end{eqnarray}
The ultraviolet divergence in (\ref{eq:aone}) may be seen by
either performing the $x$ integral first so that the resulting
$k$ integral is divergent from the high momentum limit for $d > 1$,
or by integrating over $k$ first so that the resulting $x$ integral
is divergent at the small $x$ end. For $d = 1$ there is an
additional infrared divergence coming from $k = 0$ or equivalently
$x = \infty$ behaviour.

In the following we will be interested in $1 + 1$ dimensional
theories. In that case there is a simple way to understand the
divergence of the geometric entropy
\cite{PRESK}.  Consider dividing space $x$ in a box of size $L$ into
two halves at $x = 0$.
The minimum wave number is $k_0 = {2\pi \over L}$. Now
construct a set of {\em non-overlapping} wave packets with wave
numbers in the range $2^j k_0 < k < 2^{j+1} k_0$ with integrers $j$ to
be specified in a moment. For each $j$ there is just {\em one} wave
packet which has support {\em both} in the region $x < 0$ as well as
$x > 0$. This one wavepacket alone contributes to the geometric
entropy an amount of order one. By conformal invariance we have an
equal contribution from each wave number range, i.e. for each allowed
value of $j$. However since there is an ultraviolet cutoff $\Lambda$
this allowed range is $1 < j < j_{max}$ where $j_{max} \sim {\rm
log}(L\Lambda)$. The total geometric entropy is given by some number
proportional to $j_{max}$, thus explaining the $d=1$ answer.

The above reasoning may be applied to a situation which is more
closely related to the calculation of the entanglement entropy in
an evolving black hole geometry
\cite{MATHUR},\cite{PRESK}. This is the entropy of entanglement
between the region $ x_1 < x < x_2$ and the region outside this one
on a one-dimensional line. The relevant part of entanglement is
now due to modes of wavelengths less than the size of the region
integrated over, i.e. $ \lambda < (x_2 - x_1)$. Applying the above
argument one gets
\begin{equation}
S_G (\{x_1,x_2 \}) \sim {\rm log} [(x_2 -x_1)\Lambda]
\label{eq:athree}
\end{equation}
There is another cutoff independent contribution to this entropy :
this comes from modes which are constant in the region $(x_1,x_2)$,
and will be unimportant for our purpose.

In an evaporating black hole geometry, the ultraviolet cutoff
in (\ref{eq:athree}) has to be multiplied with the scale factor
of the metric at the horizon. This gives the true significance
of the ultraviolet divergence : the evolution of the scale
factor gives the time dependence of the geometric entropy. In
this case this implies that the geometric entropy may increase
indefinitely and one may dump arbitrarily large amounts of
information into the black hole. Thus, as long as the
semiclassical approximation is valid information will be
lost.

String theories are ultraviolet finite and one may think that the
geometric entropy in string theory is finite (see Susskind in
\cite{THOFT}).  However this is not so in string perturbation
theory. It turns out that the genus one contribution is still
divergent - though this divergence may be now interpreted as an
infrared divergence related to the Hagedorn transition at the very
high local temperatures at the horizon
\cite{BARB}. String physics beyond the Hagedorn
temperature is necessarily nonperturbative in
nature.

Since two dimensional string theory is the only model for which
a nonperturbative formulation is known it is interesting to ask
how nonperaturbative effects alter the behviour of the geometric
entropy in this theory. Unfortunately despite several attempts
we still do not have a description of black holes in the matrix
model. However the notion of geometric entropy can be formulated
in the usual vacuum of the matrix model : and this is precisely
what we will do.

To understand the quantity we want to compute it is best to use the
collective field description of the fermionic field theory of the two
dimensional string \cite{DASJEV}. The dynamics of fluctuations
of the collective field is denoted by $\xi (\tau,t)$ and is
governed by the hamiltonian
\begin{eqnarray}
H_c & = & \int d\tau [ {1\over 2} (\pi_\xi^2 +
({\partial_\tau} \xi)^2) \nonumber \\
& & + {1 \over 6\rho_0^2}(({\partial_\tau} \xi)^3 +
3 \pi_\xi({\partial_\tau} \xi)\pi_\xi)]
\label{eq:bone}
\end{eqnarray}
plus some singular terms which are responsible for a {\em finite}
quantum ground state energy, but unimportant for our purposes. In
(\ref{eq:bone}) $\rho_0 (x) = {\sqrt{x^2 - \kappa}}$  is the
classical value of the collective field and we have introduced a
time of flight coordinate $\tau = \int^x dx/\rho_0$. In terms of
the space coordinate $\tau$ the effective coupling of the theory
is given by $g_{eff} = {1 \over x^2 - \kappa} = {1 \over
\kappa \sinh^2\tau}$.
The fermi level $\kappa$ therefore controls the strength of the
coupling. The coupling is weak in the asympototic region
$\tau = \pm \infty$.

In the asymptotic region the massless field $\xi$ may be related
to the massless ``tachyon'' $S(\phi,t)$ of the two dimensional string by
a nonlocal transform
\begin{equation}
S(\phi,t) = \int dk~d\tau e^{-ik(\phi-\tau)}
{\Gamma (-ik) \over \Gamma (ik)} \xi (\tau,t)
\label{eq:btwo}
\end{equation}

What we mean by the geometric entropy in this model is the following.
Pick some region ${\cal I}$ in the space defined by the
coordinate $\tau$ (or equivalently $x$ which is {\em locally} related
to $\tau$) and obtain the entropy of entanglement in the ground state
between this region and its complement. It is clear from (\ref{eq:btwo})
that this is not the same as the geometric entropy of some region in
$\phi$ space. However, we we interested in the ultraviolet behavior of the
entropy of entanglement and as we will explain later, we expect this
to be similar in $\phi$ space.

\subsection{Free nonrelativistic fermions in a box}

As explained above the geometric entropy obtained
by integrating out the degrees of freedom in half of the space,
$x < 0$ is the same as the ordinary thermodynamic entropy in a
position dependent temperature $T(x) = {1 \over 2\pi x}$. Let us
first consider the entropy {\em density} $s$ for
a gas of free nonrelativistic fermions in a box of size $L$
at temperature $T = {1 \over \beta}$ and chemical potential
$\mu_F$. The fermi momentum is $k_F = {2\pi N \over L}$.
For $\beta \mu_F >> 1$, i.e low temperatures one has
\begin{equation}
s = {\pi \over 6 \beta k_F} + {8 \pi^3 \over 45 \beta^3 k_F^5}
+ \cdots
\label{eq:bthree}
\end{equation}
The first term is the same as that for a relativistic boson
or a chiral relativistic fermion and would lead to a logarithmically
divergent geometric entropy. However the divergence comes from
the high temperature behavior where the expression (\ref{eq:bthree})
is invalid.
In this regime $\beta k_F << 1$ and one has
\begin{equation}
s = {k_F \over 2\pi}[1 - {1\over 2} {\rm ln}~({\beta k_F^2 \over 2 \pi})
+ \cdots]
\label{eq:bfour}
\end{equation}
Thus at high temperatures, the entropy density increases only
logarithmically - and this does not lead to any divergence from
the $x = 0$ end of the integral $\int_0^\infty dx s(\beta=2\pi x)$
for the geometric entropy. The $x = \infty$ behavior of this
integral is still governed by (\ref{eq:bthree}) and leads to
the standard infrared divergence. However there is no ultraviolet
divergence.

It is clear that the answer is ultraviolet finite because the
fermi sea has a finite depth. The low temperature expansion is
an expansion around the relativistic limit, in which the
depth of the fermi sea is infinite.
In terms of the collective field theory the ultraviolet
finiteness is a nonperturbative effect since for free
fermions the effective coupling is $g_{eff} = {1 \over
\mu_F}$. Perturbation theory would correspond to an expansion
around $\mu_F = \infty$ and would lead to a divergence by
virtue of (\ref{eq:bthree})

We now outline the direct calculation of the geometric entropy
from the ground state wave functional, where the above points
can be clearly seen. We will write the wave functional in
a coherent state basis in terms of grassmann fields $\psi (k)$
and ${\bar \psi} (k)$ which are eigenvalues of the corresponding
operators in appropriate coherent states. It is convenient
to shift the origin of momentum and define $k = k_F + q$ and
new fields $\chi (q) = \psi (k_F + q)$. The ground state
wave functional is then given by
\begin{eqnarray}
\Psi_0 & = & {\rm exp}~ [ -{1\over 2} \int_{-\infty}^\infty
dq~{\bar \chi} (q)\chi (q) \nonumber \\
& & + \int_{-2 k_F}^0 dq~{\bar \chi} (q) \chi (q)]
\label{eq:bfive}
\end{eqnarray}
When the terms in the exponent are converted into position space
integrals, the first term is the integral over a {\em local}
quantity. Therefore this cannot contribute to the geometric
entropy. The entire contribution comes because of the second
term, which is the contribution of the single particle modes
in the fermi sea. For a finite box size, the number of these
modes is {\em finite}. The perturbation expansion in the
collective field theory is however an expansion around the
point $k_F = \infty$. With an ultraviolet cutoff
$ \Lambda$, this means
an expansion around $k_F = \Lambda$. In fact for $k_F = \infty$
the wave functional (\ref{eq:bfive}) exactly reproduces that
for relativistic fermions with the fermi sea replaced by the
Dirac sea. In this limit, there are an large number of modes
which contribute to the geometric entropy (the largeness
controlled by the ultraviolet cutoff) and one has the usual
dependence on $\Lambda$. Non-perturbatively, however, the
finiteness of $k_F$ (which is always much less than the
ultraviolet cutoff) renders the answer independent of $\Lambda$.

To obtain an expression for the geometric entropy we have to
first expand the field in terms of modes which are localized
either in the region $ x < 0$ or in the region $ x > 0 $. A
convenient way is to write
\begin{eqnarray}
\chi (x) & = & {1 \over {\sqrt{\vert x \vert}}}
\int_{-\infty}^{\infty}{d \omega \over 2\pi}
 [ \theta (x) (x/a)^{-i\omega}
f_R (\omega) \nonumber \\
& & + \theta (-x) (- x/a)^{-i\omega} f_L (\omega)]
\label{eq:bsix}
\end{eqnarray}
The modes $f_R (\omega)$($f_L (\omega)$) are now modes localized in
the regions $x > 0$($x < 0$). One has to now express the modes $\chi
(q)$ above in terms of $f_L$ and $f_R$ to write the wave functional as
$\Psi_0 [f_L,f_R]$. The density matrix for the region $x > 0$ is then
simply given by a grassmann integration over $f_L$, from which one can
obtain the geometric entropy.  This calculation, described in
\cite{DASENT} clearly shows that geometric entropy has no dependence
on the ultraviolet cutoff.

\subsection{Geometric Entropy in Matrix Model}

The above exercise contains almost all the physics involved in
understanding the geometric entropy in the matrix model. Let us
first examine the high temperature behaviour of the theory.
Fortunately, it is sufficient to know the singlet sector
thermodynamics if we are interested in the geometric entropy in
the ground state, which is a singlet. This follows from the
relation between geometric entropy and thermodynamic entropy
with a position dependent temperature. The singlet sector
thermodynamics has been completely solved in \cite{GKLEB}
whose results will be used below.

Let $g_c$ be the critical value of the coupling in the matrix model
and define $\Delta = g_c - g$. Then the chemical potential $\mu$ is
determined by the equation
\begin{equation}
{\partial \Delta \over \partial \mu}
= {\rm Re}[\int_0^\infty dt
{(1/\kappa ) e^{-it}\over 2\pi~ {\rm sinh} (t/\kappa )}
{(\pi t/\kappa \beta)\over {\rm sinh} (\pi t/\kappa\beta)}]
\label{eq:bseven}
\end{equation}
plus an unimportant constant. The free energy is then given by
${\partial F /\partial \Delta} = \lambda^2 (\mu - \mu_c)$ and
the entropy is given by
$S = \beta^2 [({\partial F / \partial \beta})_\mu - \lambda^2\mu
({\partial \Delta /\partial \beta})_\mu]$.

The expression (\ref{eq:bseven}) has an important symmetry - $T$
duality - under $\beta \rightarrow {\pi^2 \over \beta}$ together with
$\lambda \rightarrow {\lambda \beta \over \pi}$.  The transformation
of $\lambda$ or equivalently the string coupling $g_s = {1 \over
\kappa}$ is the usual shift of the dilaton field required in
$T$-duality.  This is a very stringy symmetry and it is valid {\rm
exactly} in this model.

The standard genus expansion consists of performing a ${1 \over
\kappa}$ expansion, which means expanding both the hyperbolic
functions in (\ref{eq:bseven}) in power series. The genus one term
in the free energy is exactly equal to the genus one Polyakov path
integral calculation and may be expressed as a modular invariant
integral over the moduli space of the torus \cite{BERKLEB}.

Modular invariance is the underlying reason for ultraviolet finiteness
of string theories. However it is clear from our result that the
geometric entropy obtained by simply using this one loop answer is
ultraviolet divergent as usual. Modular invariance is clearly not
sufficient to make the geometric entropy finite in string theory. The
issue is the behaviour at very high temperatures.

Strings in two dimensions do not have a hagedorn transition.  In the
matrix model the singlet sector free energy is perfectly well defined
in the genus expansion.  This may seem to indicate that
there is nothing special about high temperatures in this carricature
string theory.

The crucial result of our work is that the genus expansion breaks down
at sufficiently high temperatures. The genus expansion \cite{GKLEB}
has the property that terms with higher and higher powers of the
string coupling contain also higher and higher powers of the
temperature - which clearly shows that while the genus expansion is
consistent with a low temperature expansion, it is inconsistent with a
high temperature expansion.

However we can use $T$-duality to obtain the high temperature bahviour
from the {\em exact} (i.e. nonperturbative) low temperature behaviour,
which can be obtained from the above formulae.  The exact low
temperature behaviour is required since $T$-duality involves a
rescaling of the string coupling $\kappa$. The results are given in
\cite{DASENT} and show that indeed there is a drastic modification of
the high temperature behaviour compared to perturbative expectations.

However these results turn out to be pathological : the specific heat
turns out to be negative in the high temperature limit. This is due to
the fact that the model of inverted harmonic oscillator with no walls
at infinity has a built-in instability in it which is, however,
invisible in the genus expansion. One may work with a regulated
potential by putting walls at a distance of order $N$ from the hump
and there cannot be any instability in this model. In fact at very
high temperatures there are a large number of very high energy fermion
excitations which perceive an almost constant potential. Using our
previous results for free fermions we then expect an ultraviolet
finite behaviour of the geometric entropy. The results would be
however necessarily non-universal.

Finally let us turn to an estimate of the geometric entropy of the
matrix model from the direct evaluation of the wave functional and the
density matrix \cite{DASMAT}.  We want to compute the entropy of
entanglement between some region $x_2 < x < x_1$ and its
complement. This quantity is logarithmically divergent (see
(\ref{eq:athree})) in 2d massless scalar field theory - which is the
one loop answer in our model. In principle, this quantity may be
computed using exact eigenstates of the inverted harmonic oscillator
hamiltonian (parabolic cylinder functions) and transforming to a basis
of modes which are localized either inside or outside this interval.

When the box $x_2 < x < x_1$ is very far away from the potential hump
and has a size $(x_1-x_2)$ much smaller than the total size, it is,
however, possible to obtain an estimate of the geometric entropy. This
is because one may use plane wave modes in this region to obtain an
estimate. The logic is similar to the one used in deriving
(\ref{eq:athree}). The entropy comes from wavepackets which are made
of waves of wavelength less than the size of the region, i.e. $\lambda
< (x_1 - x_2)$.  However for fermions the entropy of entanglement comes
{\em only} from the modes which are in the filled fermi sea. If $x_0 =
{1\over 2}(x_1+x_2)$ is the central position of the box, the effective
depth of the fermi sea at this point is ${\sqrt{x_0^2 - \kappa}} $ so
that the relevant wavelengths lie in the interval ${\sqrt{x_0^2 -
\kappa}}< \lambda < (x_1-x_2)$.  Using the expression for the
effective coupling $g_{eff}$ we get \cite{DASMAT}
\begin{equation}
S_{geom} \sim {\rm log}~[{(x_1 - x_2)^2 ~ g_{eff}(x_0)}]
\label{eq:gtwenty}
\end{equation}
The answer depends on the logarithm of the effective coupling which
now replaces the ultraviolet cutoff.

\section{Conclusions}

We have described two effects which throw light on the nature of true
degrees of freedom in two dimensional string theory. The first
demonstrates that the classical limit of the theory is rather
nontrivial : one needs more variables at the classical level than in
the full quantum theory. These extra dynamical variables are in fact
{\em quantum} fluctuations which, however, survive in the $\hbar
\rightarrow 0$ limit.  The second effect implies that the degrees of
freedom which contribute to the geometric entropy in the model are
much less in number than what is expected from perturbative
considerations and leads to a ultraviolet finite answer - the
underlying reason is the finite depth of the fermi sea and hence
nonperturbative.

At present the formalism of string theories is inadequate to determine
whether there are similar effects in higher dimensions. However, the
first effect we described has an uncanny resemblance with the
discovery that quantum effects are not suppressed even in the $g_{st}
\rightarrow 0$ limit of Type II strings moving on a Calabi Yau
conifold \cite{STROM} - though the mechanism appears to be rather
different. Finally there are several circumstancial evidences that
even higher dimensional string theories are fundamentally described by
fewer degrees of freedom than what one might naively think
\cite{HOLOGRAM}. However we dont know what these fundamental degrees
of freedom (like fermions for two dimensions) are.

\section{Acknowledgements}

I would like to thank I.C.T.P for hospitality under the Associateship
program during the Trieste Spring Workshop and the organizers of the
VIIth Regional Conference on Mathematical Physics.  I thank
S. Govindarajan, A. Jevicki, M. Li, G. Mandal, P. Mende, P. Weigmann,
S. Wadia and B. Zweibach for discussions. Finally I would like to
thank Samir Mathur for a very enjoyable collaboration and many
discussions.

\end{document}